\documentclass[print]{revtex4}
\usepackage{amssymb}
\usepackage{amscd}
\usepackage{amsfonts}
\usepackage{mathrsfs}
\usepackage{amsmath,amssymb}
\usepackage{graphicx}
\usepackage{verbatim}

\draft

\begin{document}

\title{\large \bf Topological structures of adiabatic phase for multi-level quantum systems}

\author{Zheng-Xin Liu$^{1}$\footnote{Electronic address: liuzx@mail.nankai.edu.cn},
 Xiao-Ting Zhou $^{1}$,  Xin Liu$^{1,2}$,
 Xiong-Jun Liu$^{3}$, Jing-Ling Chen$^{1}$}
\affiliation{ 1.Theoretical Physics Division, Chern Institute of
Mathematics,Nankai University, Tianjin 300071, P.R.China\\
2. Department of Physics, Texas A\&M University, College Station,
Texas 77843-4242, USA\\
3. Department of Physics, National University of Singapore, 2
Science Drive 3, Singapore 117542}

\begin{abstract}

\ \ \ \ \ \ \ \ \ \ \ \ \ \ \ \ \ \ \ \ \ \ \ \ \ \ \ \ \ \ \ \ \ \ \ \ \ \ \ \ \ \ \ \ (Dated: January 11, 2007)\\

The topological properties of adiabatic gauge fields for multi-level
(three-level in particular) quantum systems are studied in detail.
Similar to the result that the adiabatic gauge field for $SU(2)$
systems (e.g. two-level quantum system or angular momentum systems,
etc) have a monopole structure, the curvature two-forms of the
adiabatic holonomies for $SU(3)$ three-level and $SU(3)$ eight-level
quantum systems are shown to have monopole-like (for all
levels) or instanton-like (for the degenerate levels) structures.\\

\end{abstract}
\baselineskip=16pt

\maketitle

\indent

\section{introduction}
In the past twenty years, geometric phase has attracted much
attention in quantum theory. Berry firstly showed that the adiabatic
phase in a two-level system (or spin-half systems) have a monopole
field strength in the parameter space \cite{1}. Soon after that,
Simon pointed out that Berry phase is the holonomy on the $U(1)$
fibre bundle formed by the parameter manifold and the eigenstates of
the Hamiltonian, and the connection on the bundle is given by the
parallel transport condition \cite{2}. Furthermore, the geometric
phase has been extended to processes which undergo non-adiabatic (AA
phase) \cite{3} or non-cyclic (Pancharatnam phase) \cite{4}
evolution. On the other hand, the non-Abelian adiabatic phase was
first discussed by Zee and Wilczek \cite{nonAbel}. The topological
properties of non-Abelian Berry phase for $SO(2n+1)$ spinor was
studied by M.G. Benedict et.al \cite{so spinor}. After that,
P\'{e}ter L\'{e}vay studied some models with SU(2) holonomy and
pointed out some of their properties \cite{Peter}. In particular, S.
Murakami et.al. discussed the $SU(2)$ holonomy of $SO(5)$ spinor and
concluded that it is described by a Yang-monopole at the degeneracy
point \cite{8}.

We notice that multi-level (level number $N\geqslant3$) quantum
systems have been widely applied to modern technology, such as
many-channel optical interferometry \cite{interferometry}, quantum
computation \cite{computation} and quantum memory with
electromagnetically induced transparency (EIT) \cite{EIT1,EIT2},
etc. The study of the holonomy of such systems will have many
interesting potential applications, e.g. geometric quantum
computation \& geometric quantum memory, and so on. Specifically,
people can take advantage of these geometric features to the aim of
quantum information processing, as the robustness of the Berry phase
can result in a resilience against some kinds of decoherence sources
\cite{decoherence}. As a result, Berry phase in three-level systems
($SU(3)$ systems) has been discussed by many authors \cite{6,7}. The
detection of geometric phase shift in a three-level system was
achieved by Barry C. Sanders et al through a three-channel optical
interferometer \cite{experiment}. Going beyond $SU(3)$, Berry phase
for general compact Lie groups was studied by E. Strahov \cite{Lie}.

The present paper aims to discover new fundamental properties of
Geometric phase, mainly in the systems with $SU(3)$ symmetry, and
hopes to open a discussion on the topology of $SU(N)$ non-Abelian
geometric phase as well as its applications to physics. As mentioned
above, geometric phase has a wide range of applications to many
fields of physics such as quantum information science and condensate
matter physics, etc. For example, Abelian geometric phase can be
used as phase gate (see e.g. \cite{phasegate}). Barry Simon also
pointed out that in TKNN's theory of Integer Quantum Hall Effect,
the topology (the first Chern number) of $U(1)$ phase in the
momentum space (the first Brillouin Zone) explains the plateau of
Hall conductance \cite{2}. On the other hand, non-Abelian Berry
phase is promising to realize fault tolerant quantum computation
\cite{FQHE}. Especially, the topological properties for non-Abelian
Berry phase may have also important physical significance, say, that
may be found in Fractional Quantum Hall Effect and Quantum Spin Hall
Effect(see, e.g. \cite{FQHE}\cite{SQHE}). As a result, we here study
the Abelian and non-Abelian Berry phase for $SU(3)$ systems with
U(2) symmetry, and find the geometric phase in such systems has
important topological structure. We therefore hope that our results
would attract attention of physicists of Quantum Information Science
or Condensed Matter Physics, and will be applied to such research
topics in future.

The development herein is outlined as follows. In section II, we
briefly review the monopole structure of $U(1)$ holonomy for $SU(2)$
system. In section III and IV, we study the adiabatic gauge fields
for a doubly degenerate three-level system. We show that the
curvature of the field for the non-degenerate level also has a
monopole structure, while the non-Abelian curvature of the field for
the degenerate level has a instanton-like structure. The topological
properties of the adiabatic gauge fields for a non-degenerate
three-level system and a special $SU(3)$ eight-level system are
discussed in section V. In the final section we conclude our results
and present some further remarks. The detailed derivation of some
results is given in the appendix .

\section{monopole structures of the Holonomy for $SU(2)$ systems}

Before dealing with $SU(3)$ systems, let us review the holonomy for
$SU(2)$ systems. Firstly, we go over the non-degenerate case. The
adiabatic phase for a spin 1/2 particle in a magnetic field was
firstly studied by Berry\cite{1} , where a monopole structure of the
adiabatic gauge field strength was discovered. Here, for a general
consideration, we envisage a particle with spin j interacting with
an external magnetic field:
\begin{eqnarray}\label{eqn:1}
H_0(B)&=&\mu B\hat{J _z} ,\ \ \ H(\mathbf{B})=\mu\mathbf{B}\cdot
\hat{\mathbf{J}}=U(\mathbf{B})H_0U(\mathbf{B})^{\dagger},\nonumber\\
U(\mathbf{B})&=&\exp\{-iJ_z\varphi\}\exp\{-iJ_y\theta\}\exp\{-iJ_z\omega\},
\end{eqnarray}
where $\mathbf{B}(t)=B(\sin\theta (t)\cos\varphi (t),\sin\theta
(t)\sin\varphi (t),\cos\theta (t)) $. $H_0$ is the Hamiltonian in
the rest frame, and the time dependent Hamiltonian $H(\mathbf{B})$
is given by an $SU(2)$ transform of $H_0$. Here we adopt Euler's
angels $\theta, \varphi, \omega$ as the parameters of the $SU(2)$
group. There are three parameters in the Hamiltonian
$H(\mathbf{B})$, namely, the magnitude of the magnetic induction $B$
($B\geqslant0$, can be seen as the radius of the parameter space)
and two direction angles $\theta$ and $\varphi$. They span a three
dimensional Euclidian space $\mathbb R^3$. For the purpose of
simplicity, we assume that the magnitude $B$ is independent of time,
then the parameter space reduce to a sphere $S^2$. We notice that
$H_0$ is invariant under the action of an $U(1)$ subgroup,
$U(1)=\exp\{-iJ_z\omega\}$, and the coset space $SU(2)/U(1)\cong
\mathbb{C}P^1\cong S^2$, which is known as the Bloch sphere, gives
the (co)adjoint orbit of $H_0$ under the action of $SU(2)$ group. In
fact, this coset space is exactly the parameter space of
$H(\mathbf{B})$ for a given $B$. For this reason, one can rewrite
eq.(1) simply: $H(\mathbf{B})=\bar U(\mathbf{B})H_0\bar
U(\mathbf{B})^{\dagger}$, where $\bar
U(\mathbf{B})=\exp\{-iJ_z\varphi\}\exp\{-iJ_y\theta\}$.

%we only need to consider the group element in the coset space
%$SU(2)/U(1)\cong \mathbb{C}P^1\cong S^2$, which is known as the
%Bloch sphere. Thus the (co)adjoint orbit of $H_0$ with given nonzero
%$B$ is a sphere $S^2$. So the parameter space can be seen as the
%collection of all the orbits. When $B=0$, the orbit becomes a point,
%which is the origin of the parameter space $\mathbb R^3$.

 The eigenstate of $H_0$ with eigenvalue $E(B)=m\mu B$ is $\psi_0 = |m\rangle$, $m=-j,-j+1,... j$,
and the instantaneous eigenstate of $H(\mathbf{B})$ with the same
eigenvalue is $\psi (\mathbf{B})=\bar U(\mathbf{B})\psi _0$. Then
$\{(\mathbf{B},\psi(\mathbf{B})|H(\mathbf{B})\psi(\mathbf{B})=E(B)\psi\mathbf{B})\}$
defines a line bundle over the parameter space. The Berry connection
one-form for this level is:
\begin{eqnarray}\label{eqn:oneform-su2}
A=\langle \psi (\mathbf{B}) | d | \psi (\mathbf{B})\rangle
 &=&\langle \psi _0 |\bar U^{\dagger} (\mathbf{B}) d \bar U(\mathbf{B})  |\psi _0
 \rangle\nonumber\\
 &=&A_\theta d\theta+A_\varphi d\varphi,
\end{eqnarray}
where $A_{\theta}=0$,\ \ $A_{\varphi}=-im\cos\theta$.
 And the curvature two-form is:
\begin{eqnarray}\label{eqn:twoform}
F=dA=\langle \psi _0 |d\bar U^{\dagger} (\mathbf{B})\wedge d \bar
U(\mathbf{B})  |\psi _0 \rangle =\frac{1}{2}\sum_{i,j=1}^2F_{ij}
dx^i\wedge dx^j,
\end{eqnarray}
where $x^1=\theta$, $x^2=\varphi$. A straightforward calculation
gives: $F_{\theta\varphi}=im\sin\theta$. The curvature (or the gauge
field strength) can also be written as a vector:
$\mathbf{F}=\frac{im\hat{\mathbf{B}}}{B^2}$. For a cyclic evolution,
the adiabatic phase  is given by:
\begin{eqnarray}
\gamma _g =i\int _{S,\partial S=C} \mathbf{F}\cdot d\mathbf{S}=i\int
_{S, \partial S=C} \mathbf{F}\cdot \hat{\mathbf{B}} B^2 d\Omega
=-m\Omega _{(S)},
\end{eqnarray}
where C is a closed path of the magnetic induction vector in the
parameter space, S is a surface with edge C, and $\Omega _{(S)}$ is
the solid angle of the surface S. The integral $\mathbf{F}$ through
a surface surrounding the origin gives the first Chern number:
\begin{eqnarray}
\frac{i}{2\pi}\oint_{S^2} \mathbf{F}\cdot
d\mathbf{S}=\frac{i}{2\pi}\oint_{S^2}
 \mathbf{F}\cdot \hat{\mathbf{B}} B^2 d\Omega =-2m.
\end{eqnarray}

 So, we can conclude that $\mathbf{F}$ describes a gauge field with
 a Dirac monopole at the origin, whose strength (i.e. the first
 Chern number) is $-2m$, where $m=-j,-j+1,...j$.

Now, we turn to the degenerate case. If the Hamiltonian is not a
linear but a quadratic combine of the $SU(2)$ generators like this
form:
\begin{eqnarray}
H_0 =(\mu B\hat{J_z})^2,\ \ \ H(\mathbf{B})=(\mu\mathbf{B}\cdot
\hat{\mathbf{J}})^2 = \bar U(\mathbf{B})H_0\bar
U(\mathbf{B})^{\dagger}.
\end{eqnarray}

Then for $H_0$ the spin state $\psi _1 =|m\rangle$ and $\psi _2
=|-m\rangle$ ($m\neq0$) have the same energy eigenvalue $m^2 \mu ^2
B^2$, that is to say, this eigenvalue is doubly degenerate. Thus in
the adiabatic approximation, the connection one-form is:
$A^{ik}=\langle\psi_i |\bar U^{\dagger}(\mathbf{B})d\bar
U(\mathbf{B}) |\psi _k \rangle$, ($i,k=m,-m$). From
Eq.(\ref{eqn:1}), it's easy to get that $A^{ik}=0 (i\neq k)$ when
$|m-(-m)|=2m>1$. So, if $m\neq 1/2$, the connection one-form for the
doubly degenerate Hilbert space is Abelian: $A=-im\sigma_z\cos\theta
d\varphi$, and it is equal to the direct sum of the connection
one-forms of the two components, which we have discussed above. If
$j$ is a half-integral, when $m=\pm1/2$, the connection one-form is
non-Abelian:
\begin{eqnarray}
 A=-\frac{i}{2}(j+\frac{1}{2})\sigma _yd\theta
+\frac{i}{2}[-\cos\theta\sigma_z +(j+\frac{1}{2})\sin\theta\sigma _x
]d\varphi.
\end{eqnarray}
and the curvature two-form is:
\begin{eqnarray} F=dA+A\wedge A=\{-\frac{i}{2}[(j+\frac{1}{2})^2
-1]\sigma _z  \}\sin\theta d\theta\wedge d\varphi,
\end{eqnarray}
which can also be written as a vector
$\mathbf{F}=-\frac{i}{2}[(j+\frac{1}{2})^2-1]\frac{\mathbf{B}}{B^3}\sigma_z$.
It is interesting that while the gauge potential is not diagonal,
the field strength is. Similar to the non-degenerate case, the first
Chern number is given by : $\frac{i}{2\pi}\oint Tr(\mathbf{F})\cdot
d\mathbf{S}=Tr(-i[(j+\frac{1}{2})^2 -1]\sigma _z)=0$. Here each
diagonal component of $\mathbf{F}$ is a Dirac monopole, yet the
total `flux'(described by the first Chern number) is zero.
Specially, when $j=\frac{1}{2}$, the field strength $\mathbf{F}$ is
zero (because in this case the Hamiltonian is a constant).

As a natural extension, the adiabatic gauge field for $SU(m)
(m\geqslant3)$ systems has been developed by many authors. The
non-degenerate holonomy for three-level system has been discussed in
\cite{6}. The authors identified the parameter space for a pure
state (or its projection space) with the coset space
$SU(3)/SU(2)$(or the coset space $SU(3)/U(2)$), and found that the
geometric phase is proportional to the generalized solid angle in
the parameter space. The adiabatic connection and Berry phase for a
doubly degenerate three-level system have also been previously
calculated \cite{7}. However, the topological structures of the
adiabatic gauge fields are still not clear. In the rest part of this
paper, we will further investigate the topological features of the
Abelian and non-Abelian adiabatic gauge fields for $SU(3)$ systems.
In the following two sections, we focus on the topological
properties of the a adiabatic gauge fields for a doubly degenerate
three-level quantum system (a special kind of $SU(3)$ system).

\section{coadjoint Orbit of the doubly degenerate Hamiltonian for $SU(3)$ system and its geometry}

For a doubly degenerate three-level system, we first explore the
structure of the coadjoint orbit (and the parameter space) of the
Hamiltonian, with which we can easily study the adiabatic gauge
fields on it.

We choose the generators of $SU(3$) as the eight Gell-Mann matrices,
namely, $\lambda _i$, $i=1,2,\ldots 8$:
\begin{displaymath}
\lambda_1= \left(\begin{array}{cccc}0 & 1 & 0 \\ 1 & 0 & 0\\ 0 & 0 &
0\end{array} \right), \lambda_2= \left(\begin{array}{cccc}0 & -i & 0
\\ i & 0 & 0\\ 0 & 0 & 0\end{array} \right),\lambda_3= \left(\begin
{array}{cccc}1 & 0 & 0 \\ 0 & -1 & 0\\ 0 & 0 &0\end{array} \right),
\lambda_4= \left(\begin{array}{cccc}0 & 0 & 1 \\ 0 & 0 & 0\\ 1 & 0 &
0\end{array} \right),
\end{displaymath}
\begin{displaymath}
\lambda_5= \left(\begin{array}{cccc}0 & 0 & -i \\ 0 & 0 & 0\\ i & 0
& 0\end{array} \right), \lambda_6= \left(\begin{array}{cccc}0 & 0 &
0
\\ 0 & 0 & 1\\ 0 & 1 & 0\end{array} \right),\lambda_7= \left(\begin
{array}{cccc}0 & 0 & 0 \\ 0 & 0 & -i\\ 0 & i &0\end{array} \right),
\lambda_8= \frac{1}{\sqrt3}\left(\begin{array}{cccc}1 & 0 & 0 \\ 0 & 1 & 0\\
0 & 0 & -2\end{array} \right).
\end{displaymath}
They obey the following rule:
\begin{eqnarray}
Tr\{\lambda_i\lambda_j\}=2\delta_{ij},\\
\ [\lambda_i, \lambda_j]=2if_{ijk}\lambda_k,
\end{eqnarray}
where $f_{ijk}$ are the structure constants:
\begin{eqnarray}\label{f}
f_{123}=1,\ \ \ \ f_{458}=f_{678}=\frac{\sqrt3}{2},\ \ \ \
f_{147}=f_{246}=f_{257}=f_{345}=f_{516}=f_{637}=\frac{1}{2}.
\end{eqnarray}
 The Hamiltonian and its eigenstates in the rest frame are
given below respectively(We assume that the fist and the second
energy level are degenerate):
\begin{eqnarray}\label{H0}
H_0 =diag\{E_1,E_1,E_3\}=\frac{2E_1 +E_3}{3}\mathbf{I}+\frac{ E_1
-E_3}{\sqrt{3}}\lambda _8,\\
|1\rangle =(1,0,0)^T , |2\rangle =(0,1,0)^T , |3\rangle =(0,0,1)^T,
\end{eqnarray}
where $\mathbf I$ is the unit matrix, we can omit this term in the
Hamiltonian because it does not contribute to the geometric phase.
So there is only one parameter left in $H_0$, namely,
$R=\frac{(E_1-E_3)}{\sqrt{3}}$. In the adiabatic approximation, $R$
is always positive or always negative (because when $R=0$ the three
eigenvalues are degenerate and the adiabatic approximation will not
be satisfied). The sign of $R$ does not influence the eigenstates.
Generally, we take it to be positive and denote it as the radius in
the parameter space, which is similar to the magnitude $B$ in
$SU(2)$ case. Thus, the Hamiltonian in the rest frame can be
rewritten as $H_0=R\lambda_8$, and the time-dependent Hamiltonian is
given by:
\begin{eqnarray}\label{SU(3) element}
&&H(\mathbf{R})=U(\mathbf{R})H_0 U(\mathbf{R})^\dagger,\nonumber\\
&&U(\mathbf{R})=e^{(i\alpha\frac{\lambda
_3}{2})}e^{(i\beta\frac{\lambda _2}{2})}e^{(i\gamma\frac{\lambda
_3}{2})}e^{(i\theta\frac{\lambda _5}{2})}e^{(ia\frac{\lambda
_3}{2})}e^{(ib\frac{\lambda _2}{2})}e^{(ic\frac{\lambda
_3}{2})}e^{(i\phi\frac{\lambda _8}{2})},\nonumber\\
&&0\leqslant\alpha, a <2\pi;\ \ 0\leqslant\theta, \beta, b \leqslant
\pi;\ \ 0\leqslant\gamma, c<4\pi;\ \
0\leqslant\phi\leqslant2\sqrt{3}\pi.
\end{eqnarray}
Here we have used the Euler's angles of $SU(3)$ introduced in
\cite{7,9}. $\mathbf{R}$ represents the group parameters of $SU(3)$.
It's easy to see that $H_0$ is invariant under the action of the
$U(2)$ subgroup generated by $\lambda _1, \lambda _2, \lambda _3$
and $\lambda _8$, i.e. this $U(2)$ group is the isotropy group of
the Hamiltonian. Therefore the coadjoint orbit of $H_0$ is the coset
space $SU(3)/U(2)\cong \mathbb{C}P^2$ (see, for example,
\cite{CP2}). In this coset space, each group element
$\bar{U}(\mathbf{R})=\exp\{i\alpha\frac{\lambda
_3}{2}\}\exp\{i\beta\frac{\lambda _2}{2}\}\exp\{i\gamma\frac{\lambda
_3}{2}\}\exp\{i\theta\frac{\lambda _5}{2}\}$ corresponds to a new
Hamiltonian $H(\mathbf{R})$. So, for a given radii R, the
Hamiltonian's parameter space is  $\mathbb{C}P^2$. Together with the
radii R, we get the five-dimensional parameter space of the
Hamiltonian. Since $\mathbb{C}P^2\subset S^7\subset \mathbb{R}^8$
\cite{Atiyah}, here we give the eight-dimensional coordinates of the
parameter space:
\begin{eqnarray}\label{coodinate}
&&H(\mathbf{R})=\sum_{i=1}^8\xi^i\lambda_i,\ \ \
\xi^i=\frac{1}{2}Tr[H(\mathbf{R})\lambda_i],\\
&&\xi^1=\frac{\sqrt3}{2}R\sin\beta\cos\alpha\sin^2\frac{\theta}{2},
\ \ \ \ \ \ \ \ \xi^2=-\frac{\sqrt3}{2}R\sin\beta\sin\alpha\sin^2\frac{\theta}{2},\nonumber\\
&&\xi^3=-\frac{\sqrt3}{2}R\cos\beta\sin^2\frac{\theta}{2},
\ \ \ \ \  \ \ \ \ \ \ \
\xi^4=\frac{\sqrt3}{2}R\sin\theta\cos\frac{\beta}{2}\cos\frac{\alpha+\gamma}{2},\nonumber\\
&&\xi^5=\frac{\sqrt3}{2}R\sin\theta\cos\frac{\beta}{2}\sin\frac{\alpha+\gamma}{2},
\ \ \ \xi^6=\frac{\sqrt3}{2}R\sin\theta\sin\frac{\beta}{2}\cos\frac{\alpha-\gamma}{2},\nonumber\\
&&\xi^7=\frac{\sqrt3}{2}R\sin\theta\sin\frac{\beta}{2}\sin\frac{\alpha-\gamma}{2},
\ \ \ \xi^8=\frac{1}{4}R(3\cos\theta+1).
\end{eqnarray}

The north pole and the `south sphere' of the manifold
$\mathbb{C}P^2$ are intrinsically important for our later
discussion. For simplicity, we consider the unite $\mathbb{C}P^2$
with $R=1$. Considering $-\frac{1}{2}\leqslant\xi^8\leqslant1$, the
north pole is the point with coordinates $\xi^8=1, \xi^i=0(i\neq8)$,
and the south sphere is described by the equations:
$\xi^8=-\frac{1}{2}$, $\xi^4=\xi^5=\xi^6=\xi^7=0$,
$(\xi^1)^2+(\xi^2)^2+(\xi^3)^2=\frac{3}{4}$. The south sphere
$S^2_s$ describes the property of the `infinity' \cite{Kahler}.

 Now we give the metric, so that we can judge the self-dual and anti-self dual forms on the $\mathbb{C}P^2$ manifold.
Since in the Euclidian space $\mathbb R^8$, the metric is an
eight-dimensional unit matrix, so:
\begin{eqnarray}
ds^2=\sum_{i=1}^8d\xi^id\xi^i=\sum_{m,n=1}^4g_{mn}dx^mdx^n,
\end{eqnarray}
where $x^1=\beta, x^2=\alpha, x^3=\gamma, x^4=\theta$ (this gives an
orientation on $\mathbb{C}P^2$), submitting (16) into (17) we get:
\begin{displaymath} (g_{mn})=
\left(\begin{array}{cccc}\frac{3}{4}\sin^2\frac{\theta}{2} & 0 & 0 &
0\\ 0 & g_{22}
& \frac{3}{16}\cos{\beta}\sin^2\theta & 0\\
 0 & \frac{3}{16}\cos{\beta}\sin^2\theta & \frac{3}{16}\sin^2\theta & 0 \\ 0 & 0 & 0 & \frac{3}{4} \\
\end{array} \right),
\end{displaymath}
where
$g_{22}=\frac{3}{16}\sin^2\theta\cos^2\beta+\frac{3}{8}\sin^2\beta(1-\cos\theta)$.
The volume of $\mathbb{C}P^2$ is given by:
\begin{eqnarray}
V=\int_0^\pi d\theta\int_0^{2\pi} d\alpha\int_0^\pi
d\beta\int_0^{4\pi} d\gamma\sqrt{\det(g)}=\frac{9\pi^2}{2}.
\end{eqnarray}
It is known that $\mathbb{C}P^2$ is a symplectic (K\"{a}hler) space.
The symplectic (K\"{a}hler) form is given by \cite{Kahler} :
\begin{eqnarray}
\eta=\frac{1}{\sqrt3}f_{ijk}\xi_id\xi_j\wedge d\xi_k,
\end{eqnarray}
which is invariant under $SU(3)$. Let us illustrate that $\eta$ is
self-dual. Using the formula (\ref{f}) and (16), one can write this
symplectic form with the 4-dimensional coordinates:
\begin{eqnarray}
\eta=\frac{1}{2}\sum_{m,n=1}^4\eta_{mn}dx^m\wedge dx^n,
\end{eqnarray}
with
\begin{eqnarray}
\eta_{12}=\frac{3}{4}\sin\beta\sin^2\frac{\theta}{2},\ \
\eta_{24}=\frac{3}{8}\cos\beta\sin\theta,\ \ %\nonumber\\
\eta_{34}=\frac{3}{8}\sin\theta,\ \ %\nonumber\\
  else=0.
\end{eqnarray}
It is easy to verify that the above form $\eta$ is self-dual:
\begin{eqnarray}
*\eta=\frac{1}{4\sqrt{\det{(g^{-1})}}}\sum_{a,b,c,d,m,n=1}^4g_{ac}g_{bd}\varepsilon^{abmn}\eta_{mn}dx^c\wedge
dx^d=\eta.
\end{eqnarray}
where $\varepsilon^{abmn}$ is the Levi-Civita symbol. Now it's easy
to see that the volume form is given by $dV=\frac{1}{2}\eta^2$ and
that $\langle\eta,\eta\rangle=9\pi^2$, where
$\langle\alpha,\beta\rangle=\int\alpha\wedge^*\beta$ is the inner
product for forms. Furthermore, the 2-forms like:
\begin{eqnarray}\Omega_2=f_{ijk}d\xi_id\xi_jA_k(\xi)\ \ \ with\ \
\ \langle\Omega_2,\eta\rangle=0
\end{eqnarray}
span the space of anti-selfdual 2-forms. These forms are also
invariant under $SU(3)$.

\section{{topology structures of the adiabatic $U(2)$ and $U(1)$ fields for three-level system}}

In the above section, we have introduced the parameter space of the
Hamiltonian in the doubly degenerate case. Now we will probe into
the features of the adiabatic gauge fields on the parameter space.
To study the evolution of the Hamiltonian and its eigen functions,
we only need to consider the following 4-parameter group elements:
\begin{eqnarray}
\bar{U}(\mathbf{R})&=&\exp\{i\alpha\frac{\lambda
_3}{2}\}\exp\{i\beta\frac{\lambda
_2}{2}\}\exp\{i\gamma\frac{\lambda _3}{2}\}\exp\{i\theta\frac{\lambda _5}{2}\},\nonumber\\
&&|j(\mathbf{R})\rangle=\bar U(\mathbf{R})|j\rangle,\nonumber\\
  0\leqslant&\alpha&<2\pi;\ \ \ 0\leqslant\beta,\ \theta\leqslant\pi ;\ \ \
  0\leqslant\gamma<4\pi;\ \ \
   j=1,2,3.
\end{eqnarray}
where $|1\rangle$ and $|2\rangle$ are degenerate and span the
eigen-space of level $E_1$, and $|3\rangle$ is the eigen-state of
level $E_3$, and $|j(\mathbf{R})\rangle$ is the jth instantaneous
eigenstate of the $H(\mathbf R)=\bar{U}(\mathbf{R})H_0
\bar{U}(\mathbf{R})^\dagger$.  Now we have two adiabatic gauge
fields corresponding to the two levels. In other words, the
parameter manifold and the corresponding eigenstates form two
complex bundles over the manifold $\mathbb CP^2$, and the structure
groups are $U(1)$ and $U(2)$ respectively. These two bundles are
associated to two principal bundles. The first one is the $U(1)$
principal bundle known as the Hopf bundle $P(\mathbb
CP^2,U(1))=S^5$, and the second one is the $U(2)$ principal bundle
which equals to the $SU(3)$ group manifold $P(\mathbb
CP^2,U(2))=SU(3)$,
\[
\begin{CD}
U(2) % @>>> SU(3) \\
 \longrightarrow  SU(3)\\
  \ \ \ \ \ \ \ \ \ \downarrow \\
  \ \ \ \ \ \ \ \ \ \mathbb{C}P^2 %@VVV \\CP^2
\end{CD}
\]
The connection one-forms of the two gauge fields are given below:
\begin{eqnarray}
A^{(E_1)ij}=&&\langle i(\mathbf R)|d|j(\mathbf R)\rangle
\nonumber\\=&&\langle
i|\bar{U}^{\dagger}(\mathbf{R})d\bar{U}(\mathbf{R})|j\rangle,\ \ \ \
\ \ \ \ \
\ \ \ \ \ \ \ \ \ \ \ \ \ \ \ i,j=1,2,\\
A^{(E_3)}=&&\langle
3|\bar{U}^{\dagger}(\mathbf{R})d\bar{U}(\mathbf{R})|3\rangle.
\end{eqnarray}
A straightforward calculating gives ($A=\sum_{m=1}^4A_mdx^m$, where
$x^1=\beta, x^2=\alpha, x^3=\gamma, x^4=\theta$):
\begin{eqnarray}\label{eqn:oneform-su3}
A^{(E_1)}_1&=&\frac{i}{4}[-\sin(\frac{\theta}{2}+\gamma)
+\sin(\frac{\theta}{2}-\gamma)]\sigma _x
+\frac{i}{4}[\cos(\frac{\theta}{2}-\gamma)
-\cos(\frac{\theta}{2}+\gamma)]\sigma _y,\nonumber\\
A^{(E_1)}_2&=&\frac{i}{4}[-\sin ^2 \frac{\theta}{2}\cos\beta
\mathbf{I}+(\cos ^2 \frac{\theta}{2}+1)\cos\beta\sigma _z
]\nonumber\\&&+\frac{i}{8}[\sin(\beta
-\frac{\theta}{2}+\gamma)+\sin(\beta
-\frac{\theta}{2}-\gamma)+\sin(\beta
+\frac{\theta}{2}+\gamma)+\sin(\beta
+\frac{\theta}{2}-\gamma)]\sigma
_x\nonumber\\&&+\frac{i}{8}[-\cos(\beta
-\frac{\theta}{2}+\gamma)+\cos(\beta
-\frac{\theta}{2}-\gamma)-\cos(\beta
+\frac{\theta}{2}+\gamma)+\cos(\beta
+\frac{\theta}{2}-\gamma)]\sigma _y,\nonumber\\
A^{(E_1)}_3&=&\frac{i}{4}[-\sin
^2\frac{\theta}{2}\mathbf{I}+(1+\cos^2\frac{\theta}{2})\sigma _z],\
\ \ \
A^{(E_1)}_4=0;\nonumber\\
A^{(E_3)}_1&=&0, \ \ \ A^{(E_3)}_2=\frac{i}{2}\cos\beta\sin ^2
\frac{\theta}{2},\ \ \ A^{(E_3)}_3=\frac{i}{2}\sin
^2\frac{\theta}{2},\ \ \ A^{(E_3)}_4=0.
\end{eqnarray}
It is interesting that the connection form for the degenerate level
$E_1$ have a $U(1)$ component (the terms with an unite matrix
$\mathbf{I}$), which is proportional to the connection form for
level $E_3$. Above connection forms are defined on $\mathbb CP^2$.
We can extend the field to the five dimensional parameter space with
the connection component along the radius $A_5=0,\ (x^5=R)$. Then
one can calculate the geometric phase by a integral of the
connections along a path in the parameter space: $\gamma_g=$P
$\exp\{\oint_cA\}$, here P represents path ordering and $c$ is a
closed path in the parameter space. Now we focus on the topological
structures of the adiabatic gauge fields. The curvature two-forms
are given by:
\begin{eqnarray}
F=dA+A\wedge A=\frac{1}{2}\sum_{m,n=1}^4F_{mn}dx^m\wedge dx^n.
\end{eqnarray}
The components of the curvatures are given below:
\begin{eqnarray}\label{eqn:twoform su3}
F^{(E_1)}_{12}&=&\frac{i}{4}[\sin\beta\sin^2\frac{\theta}{2}(\mathbf{I}-\sigma_z)
+\cos\beta\cos\frac{\theta}{2}\sin
^2\frac{\theta}{2}(\cos\gamma\sigma _x+\sin\gamma\sigma
_y)],\nonumber\\
F^{(E_1)}_{13}&=&\frac{i}{4}\sin^2\frac{\theta}{2}\cos\frac{\theta}{2}[\cos\gamma\sigma_x
+\sin\gamma\sigma_y],\nonumber\\
F^{(E_1)}_{14}&=&\frac{i}{4}\sin\frac{\theta}{2}[\cos\gamma\sigma_y-\sin\gamma\sigma_x],\nonumber\\
F^{(E_1)}_{23}&=&\frac{i}{4}\sin\beta\sin
^2\frac{\theta}{2}\cos\frac{\theta}{2}
[-\cos\gamma\sigma_y+\sin\gamma\sigma_x],\nonumber\\
F^{(E_1)}_{24}&=&\frac{i}{4}[\frac{1}{2}\cos\beta\sin\theta(\mathbf{I}+\sigma
_z)+\sin\beta\sin\frac{\theta}{2}(\cos\gamma\sigma
_x+\sin\gamma\sigma_y)],\nonumber\\
F^{(E_1)}_{34}&=&\frac{i}{8}\sin\theta[\mathbf{I}+\sigma_z];\nonumber\\
F^{(E_3)}_{12}&=&-\frac{i}{2}\sin\beta\sin^2\frac{\theta}{2},\ \ \
F^{(E_3)}_{24}=-\frac{i}{4}\cos\beta\sin\theta,\nonumber\\
F^{(E_3)}_{34}&=&-\frac{i}{4}\sin\theta,\ \ \ \ \ \ \ \ \ \ \ \
F^{(E_3)}_{23}=F^{(E_3)}_{13}=F^{(E_3)}_{14}=0.
\end{eqnarray}
Now we can see that the two-form $F^{(E1)}$ can be decomposed into
an $SU(2)$ part and an $U(1)$ part
$F^{(E_1)}=F^{E_1[U(1)]}+F^{E_1[SU(2)]}$. The $U(1)$ part
$F^{E_1[U(1)]}$ together with the two-form $F^{(E_3)}$ is
proportional to the self-dual form given in Eq.(20) and Eq.(21):
$F^{E_1[U(1)]}=\frac{i}{3}\mathbf{I}\eta$,
$F^{(E_3)}=-\frac{2i}{3}\eta$, whereas the $SU(2)$ part
$F^{E_1[SU(2)]}$ satisfies Eq.(23) and is anti-selfdual.

The de Rham cohomology of $\mathbb CP^2$ is given by
$H^2(\mathbb{C}P^2)=\mathbb{R}\eta$ and
$H^4(\mathbb{C}P^2)=\mathbb{R}\eta^2$. The integer cohomology
$H^{2*}(\mathbb{C}P^2;\mathbb{Z})$ is generated by
$\omega=\frac{\eta}{3\pi}$ \cite{Kahler}, i.e.
\begin{eqnarray}
\int_{S^2_s}\omega=\int_{CP^2}\omega\wedge\omega=1,
\end{eqnarray}
where the 2- and 4- cycles are represented by the south sphere
$S^2_s$ and $\mathbb CP^2$ itself. This nontrivial topological
property of $\mathbb CP^2$ lead to the nontrivial properties of the
gauge fields on it. For bundles over $\mathbb CP^2$, the first and
the second Chern classes are well defined: $c_1=\frac{i}{2\pi}TrF$,
and $c_2=\frac{1}{8\pi^2}[Tr(F\wedge F)-TrF\wedge Tr F]$. And the
Chern numbers are given by:
\begin{eqnarray}
c_1&=&\frac{i}{2\pi}\int_{S^2_s}TrF, %=\frac{i}{2\pi}\int_{\beta=0}^{\pi}\int_{\alpha=0}^{2\pi}TrF,
\nonumber\\
c_2&=&\frac{1}{8\pi^2}\int_{CP^2}[Tr(F\wedge F)-TrF\wedge Tr
F].
%\nonumber\\&=&\frac{1}{8\pi^2}\int_{\beta=0}^{\pi}\int_{\alpha=0}^{2\pi}
%\int_{\gamma=0}^{\pi}\int_{\theta=0}^{\pi}[Tr(F\wedge F)-TrF\wedge Tr F].
\end{eqnarray}
One can easily calculate the Chern numbers for the bundles which we
have discussed:
\begin{eqnarray}
c_1(E_3)=1,\ \  c_1(E_1)=-1,\ \  c_2(E_1)=1.
\end{eqnarray}
This indicates that the $U(1)$ gauge field (for level $E(3)$) has a
monopole with `charge' 1, and the $U(2)$ gauge field (for level
$E_3$) has a instanton-like structure. The latter is not a usual
instanton, because the field strength is neither self-dual nor
anti-selfdual. However, just like the usual instanton, the 'action'
$I^{(E_3)}$ of the $U(2)$ gauge field is coincident with the second
Chern number $c_2$ \cite{Kahler}:
\begin{eqnarray}
c_2&=&\frac{1}{8\pi^2}\int_{CP^2}[Tr(F^{(E_3)}\wedge F^{(E_3)})-Tr
F^{(E_3)}\wedge Tr F^{(E_3)}]
\nonumber\\&=&-\frac{1}{8\pi^2}\int_{CP^2}Tr[F^{(E_3)}\wedge_*F^{(E_3)}]\nonumber\\
&\propto& I^{(E_3)}.
\end{eqnarray}
It is known that $\mathbb CP^2$ is not a spin manifold but a spinc
manifold. There is no global $SU(2)$ spinor section on $\mathbb
CP^2$, but here we have induced a bundle whose structure group is
$U(2)$. It is interesting that this bundle has the properties of
both monopole (described by $c_1$) and instanton (described by
$c_2$).

\section{adiabatic gauge fields for the completely non-degenerate three-level system
and the degenerate $SU(3)$ eight-level system}

Above we have discussed the two-fold degenerate system. For a
non-degenerate three-level system, the Hamiltonian have a symmetry
of $U(1)\otimes U(1)$(generated by $\lambda_3$ and $\lambda_8$), and
the parameter space is given by the coset $SU(3)/U(1)\otimes
U(1)=F_2$, which is a flag space. This flag space is topologically a
$S^2$ bundle over $\mathbb CP^2$ \cite{flag}. The group elements in
the coset space and the Hamiltonian are given by:
\begin{eqnarray}
&&H_0=R_3\lambda_3+R_8\lambda_8,\ \ \ \ \ \ \ H(\mathbf
R)=\widetilde{U}(\mathbf R)H_0\widetilde{U}^\dagger(\mathbf
R),\nonumber\\ &&\widetilde{U}(\mathbf R)=\bar U(\mathbf
R)\exp\{ia\frac{\lambda_3}{2}\}\exp\{ib\frac{\lambda_2}{2}\}, \ \ \
\ ( 0\leqslant b\leqslant\pi,\ \ 0\leqslant a<2\pi).
\end{eqnarray}
where $\bar U(\mathbf R)$ is given by Eq.(24). The adiabatic gauge
fields are all $U(1)$ fields. There is some discussion about the
property of these fields in \cite{6}, here we reinspect it from
another point of view. In fact, the three eigenstates are
equivalent, that is to say, we can change the order of the three
eigen values of $H_0$ via a similarity transformation. Therefore, we
only need to consider the adiabatic field for the third eigenstate
$(0,0,1)^T$. Since the projective space for a three-dimensional
Hilbert space is just $\mathbb CP^2$, the parameter space for the
Hamiltonian gives redundant information for its eigenstate (the
angles $a$ and $b$ in $\widetilde{U}(\mathbf R)$ are redundant),
thus the components of the gauge potential components along these
variables are zero. The nonzero components of the adiabatic gauge
potential are the same as $A^{(E_3)}$ [see Eq.(27)], and the
curvature two-form $F$ is the same as $F^{(E_3)}$ [see Eq.(29)].

Since the the second Betti number (the dimension of the second
cohomology group) of the flag space $F_2$ is $b_2=2$ \cite{coset},
and we can easily find two closed two-forms on $F_2$: one is $\eta$
which is defined by equations (20)\&(21), the other is $\tau=\sin
bdb\wedge da$, which is the volume form on the fiber $S^2$. so the
second de Rham cohomology of $F_2$ can be written as:
$H^2(F_2)=\mathbb R\eta+\mathbb R \tau$. Since the parameters on the
fiber $S^2$ is redundant, so the two-form $\tau$ doesn't give a
topological number for the bundles over $F_2$. And the self-dual
form $\eta$ corresponds to the fist Chern number
($F=F^{(E_3)}=-\frac{2i}{3}\eta$):
\begin{eqnarray}
c_1=\frac{i}{2\pi}\oint_{S^2_s}F=1.
\end{eqnarray}
So the $U(1)$ monopole defined on $\mathbb CP^2$ in the degenerate
case also exists on the parameter space in the non-degenerate case.

Above we have discussed the degenerate and non-degenerate $SU(3)$
three-level systems. Since the $U(1)$ monopole of $SU(2)$ system can
be extended to high-dimensional representation of
 $SU(2)$, the fibre bundles for $SU(3)$ systems can also be extended to
high-dimensional representation of $SU(3)$. As an example, we study
here the adjoint representation of $SU(3)$, which describes an
eight-level system. For this we chose the generators as
$(\Lambda_i)_{jk}=if_{ijk}$, where $f_{ijk}$ is given by
Eq.(\ref{f}). Since $\Lambda_3$ and $\Lambda_8$ are not diagonal, we
can diagonalize them simultaneously by a unitary transformation:
$\Lambda^{'}_i=V^\dagger\Lambda_iV$ (See the appendix).

Since the geometric phase for a general $SU(3)$ eight-level system
has no new topological structure, we only consider the Hamiltonian
with an $U(2)$ symmetry (then the eigenvalues have an extra
degeneracy). Similar to Eq.(12), the Hamiltonian in the rest frame
is $H_0=R\Lambda^{'}_8$. And the time dependent Hamiltonian is given
by:
\begin{eqnarray}
H(\mathbf{R})&=&\bar U(\mathbf{R}) H_0\bar U(\mathbf{R})^\dagger,
\nonumber\\\bar U(\mathbf{R})&=&\exp\{i\alpha\frac{\Lambda^{'}
_3}{2}\}\exp\{i\beta\frac{\Lambda^{'}
_2}{2}\}\exp\{i\gamma\frac{\Lambda^{'}_3}{2}\}\exp\{i\theta\frac{\Lambda^{'}_5}{2}\}.
\end{eqnarray}
The parameter space of the Hamiltonian for a given $R$ is also
$\mathbb CP^2$. The Hamiltonian have three eigenvalues:
$\pm\frac{\sqrt{3}}{2}R$ and $0$. The former two levels are both
doubly degenerate, and the third one is four-fold degenerate. Now we
have two $U(2)$ adiabatic fields and a quasi-$U(4)$ gauge field
(because the field strength of the four-fold degenerate space is
traceless and reducible). The gauge potential and the strength of
the fields are given in the appendix . The first Chern number and
second Chern number is given by (the marks $\pm$ and $0$ symbol the
energy levels):
\begin{eqnarray}
c_1^{(-)}&=&\frac{i}{2\pi}\int_{S^2_s}TrF^{(-)}=3, \nonumber\\
c_1^{(0)}&=&\frac{i}{2\pi}\int_{S^2_s}TrF^{(0)}=0, \nonumber\\
c_1^{(+)}&=&\frac{i}{2\pi}\int_{S^2_s}TrF^{(+)}=-3;\nonumber\\
c_2^{(-)}&=&\frac{1}{8\pi^2}\int_{CP^2}[Tr(F^{(-)}\wedge F^{(-)})-TrF^{(-)}\wedge TrF^{(-)}]=3, \nonumber\\
c_2^{(0)}&=&\frac{i}{8\pi^2}\int_{CP^2}[Tr(F^{(0)}\wedge F^{(0)})-TrF^{(0)}\wedge TrF^{(0)}]=3, \nonumber\\
c_2^{(+)}&=&\frac{i}{8\pi^2}\int_{CP^2}[Tr(F^{(+)}\wedge
F^{(+)})-TrF^{(+)}\wedge TrF^{(+)}]=3.
\end{eqnarray}

So, for the adiabatic $U(2)$ fields with level
$\pm\frac{\sqrt{3}}{2}R$, we have got similar results to our former
discussion. However, for level $0$, it is very interesting that the
strength of the four dimensional gauge field can be divided into an
$U(1)$ field whose field strength is zero and a three-dimensional
$SU(2)$ gauge field (or a spin-1 $SU(2)$ gauge field). This $SU(2)$
gauge filed is anti-self dual with finite 'Yang-Mills Action', which
is described by the second Chern number. So we can say that we have
got an instanton on $\mathbb CP^2$. Remembering that the parameter
space is a five-dimensional manifold spanned by the radii $R$ and a
four-dimensional 'surface' $\mathbb CP^2$, analogous to the $SU(2)$
Yang-monopole gauge field on $S^4$ or $\mathbb{R}^5$ (which is
described by the second Chern number) \cite{Yang}\cite{ 8}, we can
also say that we have got a Yang-monopole-like gauge field on the
parameter space. Actually, this result is based on the fact that
$H^{4}(\mathbb{C}P^2;\mathbb{Z})$ is nonzero, which leads to the
nonzero second Chern number.

\section{Conclusions and Further discussions}

In conclusion, we have discussed the topology structure of the
adiabatic gauge field in $SU(3)$ quantum systems. For the
twofold-degenerate three-level system (the Hamiltonian has an $U(2)$
symmetry), we find that on the parameter space ($\mathbb CP^2$ or
the five-dimensional space), the curvature two-form for the
non-degenerate level has the feature of $U(1)$ monopole, while the
curvature for the degenerate level has an either instanton-like or
monopole-like structure. The adiabatic gauge fields for a
non-degenerate three-level system have also been discussed and are
shown to have monopole structures too. It is interesting that for
the $SU(3)$ adjoint representation system (eight-level system) whose
Hamiltonian has an $U(2)$ symmetry, the adiabatic gauge field for
the four-fold-degenerate level has a Yang-monopole-like structure.
We expect that our results would attract attention of physicists of
Quantum Information Science or Condensed Matter Physics, and will be
applied to such research topics in future.

It will be interesting to generalize the present results to general
multi-level system, say, $SU(N) (N>3)$ systems. In the defining
representation, when the Hamiltonian for $SU(N)$ system has an
$U(N-1)$ symmetry, we get an $U(N-1)$ gauge field and an $U(1)$
field on the parameter space $\mathbb CP^{(N-1)}\cong SU(N)/U(N-1)$,
one can expect that the former may have an instanton-like structure
while the latter may have a monopole-like structure. General
$N$-level quantum systems have been employed in many applications to
quantum information processing and quantum information storage
\cite{multi-level}. Therefore, to study the geometric phase and
topological properties in these systems will be an interesting
issue, and will be involved in our next investigation.

We thank Prof. Mo-Lin Ge, Prof. Yong-Shi Wu, Dr. Yong Zhang ,
comrade Long-Min Wang and Dr. Suo Jin for useful discussions. This
work is supported by NSF of China under grants No.10275036, NUS
academic research Grant No. WBS:
 R -144-000-172-101 and US NSF under the grant
DMR-0547875.

\section{appendix}

The generators of the adjoint representation of $SU(3)$ are given by
$(\Lambda_i)_{jk}=if_{ijk}$. Since $\Lambda_3$ and $\Lambda_8$ are
not diagonal, we can diagonalize them by a matrix $V$ with
$\Lambda^{'}_3=V^\dagger\Lambda_3V$,
$\Lambda^{'}_8=V^\dagger\Lambda_8V$:
\begin{displaymath}
 V=\left(\begin{array}{cccccccc}
         0  &         0  &       i/\sqrt2  &\ \ \    0  & \ \ \ \   i/\sqrt2&\ \ \   0  & \ \ \        0  &         0  \\
         0  &         0  &       -1/\sqrt2 &\ \ \    0  & \ \ \ \    1/\sqrt2&\ \ \  0  &\ \ \         0  &         0  \\
         0  &         0  &         0  &\ \ \         1  &\ \ \ \      0  &\ \ \      0  &\ \ \         0  &         0  \\
      i/\sqrt2&       0  &         0  &\ \ \         0  &\ \ \ \      0  &\ \ \      0  &\ \ \      i/\sqrt2&       0  \\
       -1/\sqrt2&     0  &         0  &\ \ \         0  &\ \ \ \      0  &\ \ \      0  &\ \ \       1/\sqrt2&      0  \\
         0  &       i/\sqrt2  &    0  &\ \ \         0  &\ \ \ \      0  &\ \ \      0  &\ \ \         0  &      i/\sqrt2  \\
         0  &       -1/\sqrt2 &    0  &\ \ \         0  &\ \ \ \      0  &\ \ \      0  &\ \ \         0  &       1/\sqrt2  \\
         0  &         0  &         0  &\ \ \         0  &\ \ \ \      0  &\ \ \      1  &\ \ \         0  &         0       \\
\end{array} \right),
\end{displaymath}
\begin{displaymath}
\Lambda^{'}_3=\left(\begin{array}{cccccccc}
-\frac{1}{2}&       0&         0&         0&         0&         0&         0&         0\\
         0&   \frac{1}{2}&     0&         0&         0&         0&         0&         0\\
         0&         0&        -1&         0&         0&         0&         0&         0\\
         0&         0&         0&         0&         0&         0&         0&         0\\
         0&         0&         0&         0&         1&         0&         0&         0\\
         0&         0&         0&         0&         0&         0&         0&         0\\
         0&         0&         0&         0&         0&         0&   \frac{1}{2}&     0\\
         0&         0&         0&         0&         0&         0&         0&      -\frac{1}{2}\\
\end{array} \right),\ \ \ \
\Lambda^{'}_8=\left(\begin{array}{cccccccc}
      -\frac{\sqrt3}{2}&         0&       0&         0&         0&         0&         0&         0\\
         0&       -\frac{\sqrt3}{2}&      0&         0&         0&         0&         0&         0\\
         0&         0&         0&         0&         0&         0&         0&         0\\
         0&         0&         0&         0&         0&         0&         0&         0\\
         0&         0&         0&         0&         0&         0&         0&         0\\
         0&         0&         0&         0&         0&         0&         0&         0\\
         0&         0&         0&         0&         0&         0&       \frac{\sqrt3}{2}&         0\\
         0&         0&         0&         0&         0&         0&         0&       \frac{\sqrt3}{2}\\
\end{array} \right),
\end{displaymath}
Other six generators $\Lambda_i^{'}=V\Lambda_iV^\dagger$ are given
below:
\begin{displaymath}
\Lambda^{'}_1=\left(\begin{array}{cccccccc}
        0&            -\frac{1}{2} &            0&                  0&                  0&                  0&                  0&                  0     \\
  -\frac{1}{2}&             0&                  0&                  0&                  0&                  0&                  0&                  0     \\
        0&                  0&                  0&           -\frac{i}{\sqrt2}&         0&                  0&                  0&                  0     \\
        0&                  0&             \frac{i}{\sqrt2}&        0&           -\frac{i}{\sqrt2}&         0&                  0&                  0     \\
        0&                  0&                  0&            \frac{i}{\sqrt2}&         0&                  0&                  0&                  0     \\
        0&                  0&                  0&                  0&                  0&                  0&                  0&                  0     \\
        0&                  0&                  0&                  0&                  0&                  0&                  0&             \frac{1}{2}\\
        0&                  0&                  0&                  0&                  0&                  0&             \frac{1}{2}&             0\\
\end{array} \right),\ \
\Lambda^{'}_2=\left(\begin{array}{cccccccc}
        0&             \frac{i}{2} &            0&                  0&                  0&                  0&                  0&                  0     \\
  -\frac{i}{2}&             0&                  0&                  0&                  0&                  0&                  0&                  0     \\
        0&                  0&                  0&           -\frac{1}{\sqrt2}&         0&                  0&                  0&                  0     \\
        0&                  0&             -\frac{1}{\sqrt2}&       0&            -\frac{1}{\sqrt2}&        0&                  0&                  0     \\
        0&                  0&                  0&        -\frac{1}{\sqrt2}&            0&                  0&                  0&                  0     \\
        0&                  0&                  0&                  0&                  0&                  0&                  0&                  0     \\
        0&                  0&                  0&                  0&                  0&                  0&                  0&             \frac{i}{2}\\
        0&                  0&                  0&                  0&                  0&                  0&             -\frac{i}{2}&            0\\
\end{array} \right),
\end{displaymath}
\begin{displaymath}
\Lambda^{'}_4=\left(\begin{array}{cccccccc}
        0&                  0&                  0&            -\frac{i}{\sqrt8}&       0&     -i\sqrt{\frac{3}{8}}&             0&             0\\
        0&                  0&                  0&                  0&           \frac{1}{2}&               0&                  0&             0\\
        0&                  0&                  0&                  0&                  0&                  0&                  0&        -\frac{1}{2}\\
  \frac{i}{\sqrt8}&         0&                  0&                  0&                  0&                  0&            -\frac{i}{\sqrt8}&   0\\
        0&             \frac{1}{2}&             0&                  0&                  0&                  0&                  0&             0\\
i\sqrt{\frac{3}{8}}&        0&                  0&                  0&                  0&                  0&         -i\sqrt{\frac{3}{8}}&   0\\
        0&                  0&                  0&            \frac{i}{\sqrt8}&         0&    i\sqrt{\frac{3}{8}}&              0&             0\\
        0&                  0&        -\frac{1}{2}&                 0&                  0&                  0&                  0&             0\\
\end{array} \right),
\end{displaymath}
\begin{displaymath}
\Lambda^{'}_5=\left(\begin{array}{cccccccc}
        0&                  0&                  0&        -\frac{1}{\sqrt8}&           0&    -\sqrt{\frac{3}{8}}&               0&             0\\
        0&                  0&                  0&                  0&            -\frac{i}{2}&             0&                  0&             0\\
        0&                  0&                  0&                  0&                  0&                  0&                  0&          \frac{i}{2}\\
  -\frac{1}{\sqrt8}&        0&                  0&                  0&                  0&                  0&            -\frac{1}{\sqrt8}&   0\\
        0&           \frac{i}{2}&               0&                  0&                  0&                  0&                  0&             0\\
-\sqrt{\frac{3}{8}}&        0&                  0&                  0&                  0&                  0&           -\sqrt{\frac{3}{8}}&  0\\
        0&                  0&                  0&        -\frac{1}{\sqrt8}&            0&     -\sqrt{\frac{3}{8}}&             0&             0\\
        0&                  0&        -\frac{i}{2}&                 0&                  0&                  0&                  0&             0\\
\end{array} \right),
\end{displaymath}
\begin{displaymath}
\Lambda^{'}_6=\left(\begin{array}{cccccccc}
        0&                  0&           \frac{1}{2}&               0&                  0&                  0&                  0&                  0\\
        0&                  0&                  0&     \frac{i}{\sqrt8}&                0&        -i\sqrt\frac{3}{8}&           0&                  0\\
 \frac{1}{2}&               0&                  0&                  0&                  0&                  0&                  0&                  0\\
        0&           -\frac{i}{\sqrt8}&         0&                  0&                  0&                  0&                  0&             \frac{i}{\sqrt8}\\
        0&                  0&                  0&                  0&                  0&                  0&           -\frac{1}{2}&              0\\
        0&          i\sqrt\frac{3}{8}&          0&                  0&                  0&                  0&                  0&            -i\sqrt\frac{3}{8}\\
        0&                  0&                  0&                  0&        -\frac{1}{2}&                 0&                  0&                  0\\
        0&                  0&                  0&      -\frac{i}{\sqrt8}&              0&       i\sqrt\frac{3}{8}&             0&                  0\\
\end{array} \right),
\end{displaymath}
\begin{displaymath}
\Lambda^{'}_7=\left(\begin{array}{cccccccc}
        0&                  0&                 -\frac{i}{2}&        0&                  0&                  0&                  0&                  0\\
        0&                  0&                  0&       \frac{1}{\sqrt8}&              0&         -\sqrt\frac{3}{8}&           0&                  0\\
       \frac{i}{2}&         0&                  0&                  0&                  0&                  0&                  0&                  0\\
        0&       \frac{1}{\sqrt8}&              0&                  0&                  0&                  0&                  0&       \frac{1}{\sqrt8}\\
        0&                  0&                  0&                  0&                  0&                  0&             \frac{i}{2}&             0\\
        0&      -\sqrt\frac{3}{8}&              0&                  0&                  0&                  0&                  0&       -\sqrt\frac{3}{8}\\
        0&                  0&                  0&                  0&         -\frac{i}{2}&                0&                  0&                  0\\
        0&                  0&                  0&       \frac{1}{\sqrt8}&              0&          -\sqrt\frac{3}{8}&          0&                  0\\
\end{array} \right).
\end{displaymath}

The following are the components of the gauge potential and the
strength of the adiabatic gauge field for the $SU(3)$ adjoint
representation system, where $+, -$ and $0$ label the levels of the
Hamiltonian.

\begin{displaymath}
 A^{(-)}_1=
 \left(\begin{array}{cc}
     0 &   -\frac{1}{4}(e^{i(\frac{\theta}{2}+\gamma)}+e^{i(-\frac{\theta}{2}+\gamma)})                       \\
        \frac{1}{4}(e^{-i(\frac{\theta}{2}+\gamma)}+e^{i(\frac{\theta}{2}-\gamma)})     & 0                 \\
\end{array}\right),
\end{displaymath}
\begin{displaymath}
A^{(-)}_2= \left(\begin{array}{cc}
        -\frac{i}{2}\cos\beta\cos\theta &   \frac{i}{2}\sin\beta\cos\frac{\theta}{2} e^{i\gamma}                       \\
        \frac{i}{2}\sin\beta\cos\frac{\theta}{2} e^{-i\gamma}   &     \frac{i}{4}\cos\beta(3-\cos\theta)                 \\
\end{array} \right),
\end{displaymath}
\begin{displaymath}
A^{(-)}_3= \left(\begin{array}{cc}
  -\frac{i}{2}\cos\theta &  0                  \\
       0 &  \frac{i}{4}(3-\cos\theta)                 \\
\end{array}\right),\ \ \ \ \ A^{(-)}_4=0,
\end{displaymath}

\begin{displaymath}
A^{(0)}_1= \left(\begin{array}{cccc}
  0            &-\frac{\sqrt2i}{2}\cos\frac{\theta}{2}e^{i\gamma} & 0 & 0                \\
  -\frac{\sqrt2i}{2}\cos\frac{\theta}{2}e^{-i\gamma}  &  0 & -\frac{\sqrt2i}{2}\cos\frac{\theta}{2}e^{i\gamma} &0   \\
0 & -\frac{\sqrt2i}{2}\cos\frac{\theta}{2}e^{-i\gamma}& 0 & 0\\
0&0 &0&0\\
\end{array}\right),
\end{displaymath}
\begin{displaymath}
A^{(0)}_2= \left(\begin{array}{cccc}
-\frac{i}{4}\cos\beta(3+\cos\theta)           &-\frac{\sqrt2}{2}\sin\beta\cos\frac{\theta}{2}e^{i\gamma} & 0 & 0                \\
\frac{\sqrt2}{2}\sin\beta\cos\frac{\theta}{2}e^{-i\gamma}  &  0 & -\frac{\sqrt2}{2}\sin\beta\cos\frac{\theta}{2}e^{i\gamma} &0  \\
0 & \frac{\sqrt2}{2}\sin\beta\cos\frac{\theta}{2}e^{-i\gamma}& \frac{i}{4}\cos\beta(3+\cos\theta) & 0\\
0&0&0&0\\
\end{array}\right),
\end{displaymath}
\begin{displaymath}
A^{(0)}_3= \left(\begin{array}{cccc}
-\frac{i}{4}(3+\cos\theta)  &0 & 0 & 0   \\
 0 &  0 & 0 & 0 \\
0 & 0 & \frac{i}{4}(3+\cos\theta) & 0\\
0&0 &0&0\\
\end{array}\right),A^{(0)}_4=0,
\end{displaymath}

\begin{eqnarray}
A^{(+)}_m=A^{*(-)}_m\nonumber,\ \ \ \ \ m=1,2,3,4
\end{eqnarray}

\begin{displaymath}
F^{(-)}_{12}= \left(\begin{array}{cc}
-i\sin\beta\sin^2\frac{\theta}{2}           &\frac{i}{8}\cos\beta\cos\frac{\theta}{2}\sin^2\frac{\theta}{2}e^{i\gamma}                  \\
\frac{i}{8}\cos\beta\cos\frac{\theta}{2}\sin^2\frac{\theta}{2}e^{-i\gamma}&-2i\sin\beta\sin^2\frac{\theta}{2}
\end{array}\right),
\end{displaymath}
\begin{displaymath}
F^{(-)}_{13}= \left(\begin{array}{cc}
       0 &   \frac{i}{8}\sin^2\frac{\theta}{2}(e^{i(\frac{\theta}{2}+\gamma)}+e^{i(-\frac{\theta}{2}+\gamma)})                       \\
        \frac{i}{8}\sin^2\frac{\theta}{2}(e^{-i(\frac{\theta}{2}+\gamma)}+e^{i(\frac{\theta}{2}-\gamma)})     & 0                 \\
\end{array} \right),
\end{displaymath}
\begin{displaymath}
F^{(-)}_{14}= \left(\begin{array}{cc}
       0 &   -\frac{1}{4}\sin\frac{\theta}{2}e^{i\gamma}                      \\
       \frac{1}{4}\sin\frac{\theta}{2}e^{-i\gamma}      & 0                 \\
\end{array} \right),
%\end{displaymath}
%\begin{displaymath}
F^{(-)}_{23}= \left(\begin{array}{cc}
       0 &   \frac{1}{4}\sin\beta\cos\frac{\theta}{2}\sin^2\frac{\theta}{2}e^{i\gamma}                      \\
 -\frac{1}{4}\sin\beta\cos\frac{\theta}{2}\sin^2\frac{\theta}{2}e^{-i\gamma}     & 0                 \\
\end{array} \right),
\end{displaymath}
\begin{displaymath}
F^{(-)}_{24}= \left(\begin{array}{cc}
 -\frac{i}{2}\cos\beta\sin\theta &   \frac{i}{4}\sin\beta\sin\frac{\theta}{2}e^{i\gamma}                      \\
  \frac{i}{4}\sin\beta\sin\frac{\theta}{2}e^{-i\gamma}     &  -\frac{i}{4}\cos\beta\sin\theta                \\
\end{array} \right),
F^{(-)}_{34}= \left(\begin{array}{cc}
 -\frac{i}{2}\sin\theta &  0                  \\
0    &  -\frac{i}{4}\sin\theta                \\
\end{array} \right),
\end{displaymath}

\begin{displaymath}
F^{(0)}_{12}=\left(\begin{array}{cccc}
\frac{i}{2}\sin\beta\sin^2\frac{\theta}{2}           &-\frac{\sqrt2}{4}\cos\beta\cos\frac{\theta}{2}\sin^2\frac{\theta}{2}e^{i\gamma} & 0 & 0                \\
\frac{\sqrt2}{4}\cos\beta\cos\frac{\theta}{2}\sin^2\frac{\theta}{2}e^{-i\gamma}  &  0 &  -\frac{\sqrt2}{4}\cos\beta\cos\frac{\theta}{2}\sin^2\frac{\theta}{2}e^{i\gamma}&0\\
0 & \frac{\sqrt2}{4}\cos\beta\cos\frac{\theta}{2}\sin^2\frac{\theta}{2}e^{-i\gamma} & -\frac{i}{2}\sin\beta\sin^2\frac{\theta}{2}& 0\\
0&0&0&0\\
\end{array}\right),
\end{displaymath}
\begin{displaymath}
F^{(0)}_{13}=\left(\begin{array}{cccc}
0           &-\frac{\sqrt2}{4}\cos\theta\sin^2\frac{\theta}{2}e^{i\gamma} & 0 & 0 \\
\frac{\sqrt2}{4}\cos\theta\sin^2\frac{\theta}{2}e^{-i\gamma}  &  0 & -\frac{\sqrt2}{4}\cos\theta\sin^2\frac{\theta}{2}e^{i\gamma}& 0 \\
0 &\frac{\sqrt2}{4}\cos\theta\sin^2\frac{\theta}{2}e^{-i\gamma} & 0 & 0\\
0&0&0&0
\end{array}\right),
\end{displaymath}
\begin{displaymath}
F^{(0)}_{14}=\left(\begin{array}{cccc}
0    &-\frac{\sqrt2i}{4}\sin\frac{\theta}{2}e^{i\gamma} & 0 & 0 \\
-\frac{\sqrt2i}{4}\sin\frac{\theta}{2}e^{-i\gamma}  &  0 &-\frac{\sqrt2i}{4}\sin\frac{\theta}{2}e^{i\gamma}& 0\\
0 & -\frac{\sqrt2i}{4}\sin\frac{\theta}{2}e^{-i\gamma}& 0 & 0\\
0&0&0&0
\end{array}\right),
\end{displaymath}
\begin{displaymath}
F^{(0)}_{23}=\left(\begin{array}{cccc}
0    &\frac{\sqrt2i}{4}\sin\beta\cos\frac{\theta}{2}\sin^2\frac{\theta}{2}e^{i\gamma} & 0 & 0   \\
\frac{\sqrt2i}{4}\sin\beta\cos\frac{\theta}{2}\sin^2\frac{\theta}{2}e^{-i\gamma}  &  0 &\frac{\sqrt2i}{4}\sin\beta\cos\frac{\theta}{2}\sin^2\frac{\theta}{2}e^{i\gamma} & 0  \\
0 &\frac{\sqrt2i}{4}\sin\beta\cos\frac{\theta}{2}\sin^2\frac{\theta}{2}e^{-i\gamma}& 0 & 0\\
0&0&0&0
\end{array}\right),
\end{displaymath}
\begin{displaymath}
F^{(0)}_{24}=\left(\begin{array}{cccc}
-\frac{i}{4}\cos\beta\sin\theta           &-\frac{\sqrt2}{4}\sin\beta\sin\frac{\theta}{2}e^{i\gamma} & 0 & 0                \\
\frac{\sqrt2}{4}\sin\beta\sin\frac{\theta}{2}e^{-i\gamma}   &  0 &-\frac{\sqrt2}{4}\sin\beta\sin\frac{\theta}{2}e^{i\gamma} & 0 \\
0 & \frac{\sqrt2}{4}\sin\beta\sin\frac{\theta}{2}e^{-i\gamma} &\frac{i}{4}\cos\beta\sin\theta & 0\\
0&0&0&0
\end{array}\right),
\end{displaymath}
\begin{displaymath}
F^{(0)}_{34}=\left(\begin{array}{cccc}
-\frac{i}{4}\sin\theta  &0 & 0 & 0 \\
0&  0 & 0 & 0  \\
0 & 0 & \frac{i}{4}\sin\theta & 0\\
0&0&0&0
\end{array}\right),
\end{displaymath}

\begin{displaymath}
F^{(+)}_{mn}=F^{*(-)}_{mn}\nonumber\ \ \ \ \ m,n=1,2,3,4.
\end{displaymath}

%%%%%%%%%%%%%%%%%%%%%%%%%%%%%%%%%%%%%%%%%%%%%

%%%%%%%%%%%%%%%%%%%%%%%%%%%%%%%%%%%%%%%%%%%%%

\bigskip

\noindent

\end{document}